\documentclass[aps,prc,showpacs,showkeys,twocolumn,superscriptaddress,groupedaddress]{revtex4-1}

\usepackage[english]{babel}
\usepackage{graphics}
\usepackage{graphicx}
\usepackage{bm}
\usepackage{verbatim}
\usepackage{amssymb}
\usepackage{braket}
\usepackage{color}
\usepackage{wrapfig}
\usepackage{pifont}
\usepackage{amsfonts}
\usepackage{mathrsfs}
\usepackage{slashed}
\usepackage{amsmath}
\usepackage{epsfig}
\usepackage{latexsym}
\usepackage{psfrag}
\usepackage{nicefrac}
\usepackage{ragged2e}
\usepackage{etoolbox}
\usepackage{tikz}
\usepackage{pgfpages}
\usepackage{dsfont}
\usepackage{mathtools}
\usepackage{bbm}

     \def\chiral4lo{$\mathrm{N}^4\mathrm{LO}$}\def\id{\mathds{1}}

\begin{document}

\noindent
\title{Inelastic nucleon-nucleus scattering from a microscopic point of view}

\author{Matteo Vorabbi$^{1}$}
\author{Michael Gennari$^{2}$}
\author{Paolo Finelli$^{3}$}
\author{Carlotta Giusti$^{4}$}
\author{Petr Navr\'{a}til$^{5,6}$}

\affiliation{$~^{1}$ Department of Physics, University of Surrey, Guildford, GU2 7XH, United Kingdom
}

\affiliation{$~^{2}$ Institut f\"ur Kernphysik and PRISMA++ Cluster of Excellence, Johannes Gutenberg-Universit\"at Mainz, 55128 Mainz, Germany
}

\affiliation{$~^{3}$Dipartimento di Fisica e Astronomia, 
Universit\`{a} degli Studi di Bologna and \\
INFN, Sezione di Bologna, Via Irnerio 46, I-40126 Bologna, Italy
}

\affiliation{$~^{4}$
INFN, Sezione di Pavia,  Via A. Bassi 6, I-27100 Pavia, Italy
}

\affiliation{$~^{5}$University of Victoria, 3800 Finnerty Road, Victoria, British Columbia V8P 5C2, Canada
}

\affiliation{$~^{6}$TRIUMF, 4004 Wesbrook Mall, Vancouver, British Columbia, V6T 2A3, Canada
}

\date{\today}


\begin{abstract} 
We apply to the nucleon-nucleus inelastic process a fully coherent microscopic multiple scattering approach. Our study addresses the complexities inherent in characterizing inelastic scattering events, offering a comprehensive theoretical model grounded in the reaction theory. The approach is based on the distorted-wave approximation and requires the knowledge of three potentials, which give the initial and final distorted wave functions and the transition potential. All of them are derived just like the microscopic optical potential for elastic nucleon-nucleus scattering we derived in previous papers of ours within the framework of the Watson multiple scattering theory and adopting the impulse approximation. The potentials are obtained by folding nonlocal \textit{ab initio} nuclear densities from the No-Core Shell Model (NCSM) with a nucleon-nucleon $t$ matrix computed with a chiral interaction consistent with the one used in the calculation of the density. The only difference in the formal expressions of the three potentials resides in the nuclear density, where we use the ground and excited state densities of the target and the transition density. By extending methods traditionally applied to elastic scattering, we incorporate the effects of inelastic transitions enabling an accurate description of the experimental differential cross section. 
The predictive power of our numerical results is benchmarked against empirical data of inelastic proton scattering off $^{12}$C, for the transition to the $2^+$ state at 4.44 MeV, in a range of projectile energies of 65-300 MeV. The generally good description of the experimental cross sections as functions of the scattering angle gives clear evidence of the reliability and robustness of a model that does not contain any free adjustable parameters. 

\end{abstract}


\maketitle


\section{Introduction}

Microscopic inelastic nucleon-nucleus ($NA$) processes involve complex interactions between a nucleon and a nucleus, where the kinetic energy of the nucleon is partially transferred to the nucleus, exciting it to a higher energy state. The study of inelastic $NA$ scattering is a fundamental aspect of nuclear physics, providing critical insights into the structure and dynamics of atomic nuclei. Traditional phenomenological approaches have yielded significant understanding, yet they inherently rely on empirical adjustments and fitting parameters, limiting their predictive power and universality \cite{refId0}. In recent years, there has been a concerted effort to develop 
\textit{ab initio} methods - approaches rooted in first principles and fundamental interactions - that promise a more rigorous and predictive framework for understanding $NA$ processes.

\textit{Ab initio} approaches \cite{RevModPhys.92.025004} in nuclear physics aim to describe nuclear phenomena starting from the interactions between individual nucleons, typically governed by Quantum Chromodynamics (QCD) or Effective Field Theories (EFTs) derived from QCD. These methods avoid the phenomenological models' reliance on experimental data for parameter fitting, instead utilizing computational techniques to solve the many-body Schrödinger equation directly for nucleonic systems. This paradigm shift has the goal to unify our understanding of nuclear reactions across a broad range of energies and isotopic compositions with a clear connection to the same underlying physics that governs fundamental particle interactions. 

In this paper we present a comprehensive examination of microscopic inelastic $NA$ processes through the lens of \textit{ab initio} nuclear theory. By employing state-of-the-art computational tools and high-fidelity nucleon-nucleon ($NN$) interactions, we aim to elucidate the mechanisms underpinning inelastic scattering events. Our approach integrates recent advancements in nuclear many-body methods, such as the No-Core Shell Model (NCSM) \cite{Navr_til_2009,Barrett:2013nh}, to provide a robust and detailed description of the nucleon-nucleus interaction.

The method we propose is rooted in the so-called Distorted-Wave (DW) approximation, that has been successful since the seminal work of Satchler (see Refs. \cite{SATCHLER1979183, Satchler:1987ne, 17293} for an extensive treatment). The DW approach is a theoretical method used to describe scattering processes in quantum mechanics, especially when interactions beyond the simple potential scattering must be included. Since in many scattering problems the interaction between particles cannot be treated as a simple {\it plane wave in/out} process (as in the Born approximation), the distorted-wave approach provides a more accurate description by including the effects of initial- and final-state interactions.

The approach requires three potentials, which give the distortion of the initial- and final-state wave functions (initial- and final-state interactions) and the transition from the initial ground state of the target nucleus to its final excited state.
The potential for the initial state, accounting for the interaction between the projectile and the target in its ground state, is the optical potential for elastic $NA$ scattering, and here we exploit our recent work where, in a series of papers \cite{Vorabbi:2015nra,Vorabbi:2017rvk,Vorabbi:2018bav,PhysRevC.97.034619,PhysRevLett.124.162501,Vorabbi:2020cgf,Vorabbi:2021kho,PhysRevC.109.034613,qxtf-5b4y}, we derived microscopic optical potentials from chiral EFTs. The optical potential is obtained, at first-order in the Watson multiple scattering theory and with adoption of the impulse approximation, by folding nonlocal \textit{ab initio} one-body nuclear densities from the NCSM with a $NN$ $t$ matrix computed with a chiral interaction consistent with the one used in the calculation of the density. The other two potentials are derived through generalization of our microscopic optical potential, following the same steps, and adopting the same approximations. The final formal expression of the three potentials is the same, the only difference is in the density matrix, where we have the one-body density of the ground and excited states of the target and the transition density.

The paper is structured as follows. In Sect. \ref{dw} we give a detailed overview of the distorted-wave formalism to inelastic $NA$ scattering and describe the computational methods and algorithms employed in our analysis. Details of the computational framework are given in the Appendix. Then we derive the three potentials required by our approach (Sect. \ref{theory}) and briefly discuss the $NN$ chiral potential used in the calculations (Sect. \ref{NNpotential}), and the calculation of the one-body density matrices (Sect. \ref{target}). 
In Sect. \ref{results} we present our numerical results and compare them with experimental data and with  the corresponding results of other models available in the literature for a benchmark. Finally, in Sect. \ref{conclusions}, we draw some conclusions, discuss the implications of our findings for future research in nuclear physics, and outline potential directions for further refinements and applications of 
\textit{ab initio} techniques.


\section{Distorted-wave formalism}
\label{dw}
In this section we introduce a distorted-wave formalism to describe the nucleon-nucleus inelastic scattering process using microscopic optical potentials 
along the theoretical approaches introduced in Refs. \cite{PhysRevC.25.1215,PhysRevC.25.1233}. We start by defining the total Hamiltonian for the entire
$(A+1)$-nucleon system as follows
\begin{equation}
H_{A+1} = H_0 + V \, ,
\end{equation}
where the operator $V$ represents the external interaction between the projectile and the target. Under the assumption of two-body forces only \cite{PhysRevC.103.024604}, $V$ can be written as
\begin{equation}
V = \sum_{i=1}^A v_{0 i} \, ,
\end{equation}
where $v_{0 i}$ represents the two-body interaction between the projectile (labeled with $``0"$) and the $i$th nucleon in the target nucleus.
The free Hamiltonian $H_0$ is given by
\begin{equation}
H_0 = h_0 + H_A \, ,
\end{equation}
where $h_0$ is the kinetic energy operator of the projectile in the center of mass and $H_A$ is the intrinsic Hamiltonian of the target that satisfies
\begin{equation}
H_A \ket{\Phi_n} = e_n \ket{\Phi_n} \, , 
\end{equation}
with the usual normalization condition
\begin{equation}
\braket{\Phi_m |\Phi_n } = \delta_{m n} \, , \hspace{1.0cm} m,n=0,1,2,\ldots \, .
\end{equation}
$\Phi_0$ is the ground state of the target with eigenvalue $e_0$, $\Phi_1$ is the first excited state of the target with eigenvalue $e_1$, and so on.
For convenience, $e_0$ is taken equal to zero. Defining with ${\bm k}$ and ${\bm k}^{\prime}$ the initial and final momenta of the projectile in the $NA$ center of mass, we obtain
\begin{equation}
\label{eq_H0}
H_0 \ket{\Phi_n \, {\bm k}} = E_n (k) \ket{\Phi_n \, {\bm k}} \, 
\end{equation}
where the full wavefunction is written as
\begin{equation}
\ket{\Phi_n \, {\bm k}} \equiv \ket{\Phi_n} \ket{\bm k} \, .
\end{equation}
The energies appearing in Eq. (\ref{eq_H0}) are defined in terms of the kinetic energy $E (k)$ (eigenvalue of $h_0$) of the incident particle
as follows
\begin{equation}
E_n (k) = E (k) + e_n \, .
\end{equation}
Here $E (k)$ is given by
\begin{align}
E (k) &= E_{proj} (k) + E_{target} (k) \nonumber \\
&= \sqrt{k^2 + m_{proj}^2} + \sqrt{k^2 + m_{targ}^2} \, ,
\end{align}
where $m_{proj}$ and $m_{targ}$ denote the masses of the projectile and target, respectively.
Additionally, we denote with $G_0$ the free many-body propagator for the $NA$ system
\begin{equation}
G_0 = \frac{1}{E - H_0 + i \epsilon} \, ,
\end{equation}
and for later convenience we denote with $\ket{\Phi}$ the ground state $\ket{\Phi_0}$, while we use $\ket{\Phi_{\ast}}$ to denote a general excited state of the target
$\ket{\Phi_n }$, with $n\neq 0$.

We introduce the following projection operators in the space of the $A$ target nucleons
\begin{align}
P &= \ket{\Phi} \bra{\Phi} \, , \\
P_{\ast} &= \ket{\Phi_{\ast}} \bra{\Phi_{\ast}} \, , \\
Q &= \id - P \, , \\
Q_{\ast} &= \id - P_{\ast} \, ,
\end{align}
where $P$ projects onto the target ground state $\ket{\Phi}$, $P_{\ast}$ projects onto one of the target excited states, while $Q$ and $Q_{\ast}$ are their respective
conjugate projectors.
They fulfill the following properties:
\begin{enumerate}
\item Completeness
\begin{align}
P + Q &= \id \, , \\
P_{\ast} + Q_{\ast} &= \id \, ,
\end{align}
\item Idempotency
\begin{align}
P^2 &= P \, , \quad \, \, Q^2 = Q \, , \\
P_{\ast}^2 &= P_{\ast} \, , \quad Q_{\ast}^2 = Q_{\ast} \, ,
\end{align}
\item Orthogonality
\begin{align}
P Q &= Q P = 0 \, , \\
P_{\ast} Q_{\ast} &= Q_{\ast} P_{\ast} = 0 \, .
 \end{align}
\end{enumerate}
By construction, the projectors $P$ and $P_{\ast}$ commute with the free projectile-target propagator
\begin{equation}
[G_0 , P] = [G_0 , P_{\ast}] = 0 \, .
\end{equation}

With all these definitions we proceed to calculate the fundamental quantity of interest for inelastic scattering, {\it i.e.} the transition matrix
element $T_{f i}^{\mathrm{inel}}$, which defines a transition from an initial state $i$ to a final state $f$.
In particular, we are interested in inelastic transitions from the ground state to an excited state of the target and, thus, we introduce the transition amplitude as
\begin{equation}\label{definition_of_transition_amplitude}
T_{f i}^{\mathrm{inel}} = \braket{{\bm k}^{\prime} \Phi_{\ast} | T | \Phi \, {\bm k}} \, ,
\end{equation}
where the transition operator $T$ satisfies the $(A+1)$-body Lippmann-Schwinger equation
\begin{equation}
T = V + V G_0 T \, ,
\end{equation}
and its formal solution is
\begin{equation}\label{formal_solution_of_general_trans_amplitude}
T = V {(\id - G_0 V)}^{-1} \, .
\end{equation}

As a first step we provide a restatement of Eq.~(\ref{definition_of_transition_amplitude}) in terms of the distorted waves.

In nuclear reaction theory, distorted waves are solutions to the Schrödinger equation that include the effect of the nuclear potential experienced by the incoming and outgoing particles. Unlike plane waves, which assume free motion, distorted waves account for the distortion of particle trajectories due to the complex interactions with the target nucleus. They are typically calculated using an optical potential that includes both real (mean field) and imaginary (absorptive) parts to model elastic scattering and reaction losses, respectively.

If we substitute Eq.~(\ref{formal_solution_of_general_trans_amplitude}) into Eq.~(\ref{definition_of_transition_amplitude}), we obtain
\begin{equation}\label{second_general_transition_amplitude}
T_{f i}^{\mathrm{inel}} = \braket{{\bm k}^{\prime} \Phi_{\ast} | V | \Psi^{(+)}} \, ,
\end{equation}
where $\ket{\Psi^{(+)}}$ satisfies the equation
\begin{equation}\label{full_state_vector}
\ket{\Psi^{(+)}} = \ket{ \Phi \, {\bm k}} + G_0 V \ket{\Psi^{(+)}} \, ,
\end{equation}
and represents the full state vector for the $NA$ system. It has the incident boundary condition of a plane wave for a projectile with momentum ${\bm k}$
impinging upon the target nucleus in its ground state $\ket{\Phi}$. 

The vector $\ket{\Psi^{(+)}}$ represents the outgoing scattering eigenfunction of the Hamiltonian $H_{A+1}$ and contains the complexity of the full many-body problem.
Acting with the operators $P$ and $Q$ on Eq.~(\ref{full_state_vector}) we obtain
\begin{equation}\label{elastic_state_vector}
P \ket{\Psi^{(+)}} = \ket{ \Phi \, {\bm k}} + G_0 P V \ket{\Psi^{(+)}} \, ,
\end{equation}
and
\begin{equation}\label{inelastic_vector_state}
Q \ket{\Psi^{(+)}} = G_0 Q V \ket{\Psi^{(+)}} \, .
\end{equation}
Using the identity $\id = P + Q$ on the state vector $\ket{\Psi^{(+)}}$, on the right-hand side of Eq.~(\ref{elastic_state_vector}) and Eq.(\ref{inelastic_vector_state}), we obtain
\begin{equation}\label{second_elastic_state_vector}
P \ket{\Psi^{(+)}} = \ket{ \Phi \, {\bm k}} + G_0 P V (P+Q) \ket{\Psi^{(+)}} \, ,
\end{equation}
and
\begin{equation}\label{second_inelastic_vector_state}
Q \ket{\Psi^{(+)}} = {(\id - G_0 Q V)}^{-1} G_0 Q V P \ket{\Psi^{(+)}} \, .
\end{equation}
Noting that
\begin{equation}
\ket{\Psi^{(+)}} = P \ket{\Psi^{(+)}} + Q \ket{\Psi^{(+)}} \, ,
\end{equation}
and using Eq.~(\ref{second_inelastic_vector_state}) we obtain
\begin{equation}\label{reexpression_of_full_vector_state}
\ket{\Psi^{(+)}} = {(\id - G_0 Q V)}^{-1} P \ket{\Psi^{(+)}} \, .
\end{equation}
By Eq.~(\ref{reexpression_of_full_vector_state}) we can rewrite the transition amplitude (\ref{second_general_transition_amplitude}) as follows
\begin{equation}\label{third_general_transition_amplitude}
T_{f i}^{\mathrm{inel}} = \braket{{\bm k}^{\prime} \Phi_{\ast} | U | \psi^{(+)}} \, ,
\end{equation}
where $U$ is an auxiliary optical potential
\begin{equation}\label{first_definition_elastic_opt_pot}
U \equiv V {(\id - G_0 Q V)}^{-1} \, ,
\end{equation}
and
\begin{equation}\label{definition_initial_dist_wave}
\ket{\psi^{(+)}} \equiv P \ket{\Psi^{(+)}} \, ,
\end{equation}
with the property $P \ket{\psi^{(+)}} = \ket{\psi^{(+)}}$.
The initial distorted wave $\ket{\psi^{(+)}}$ can be obtained by inserting
Eq.~(\ref{second_inelastic_vector_state}) into Eq.~(\ref{second_elastic_state_vector}) and using
 Eqs.~(\ref{first_definition_elastic_opt_pot}) and
(\ref{definition_initial_dist_wave}) 
\begin{equation}\label{initial_distorted_wave}
\ket{\psi^{(+)}} = \ket{\Phi \, {\bm k}} + G_0 P U P \ket{\psi^{(+)}} \, .
\end{equation}
We also note from Eq.~(\ref{first_definition_elastic_opt_pot}) that $U$ can be rewritten as
\begin{equation}\label{second_definition_elastic_opt_pot}
U = {(\id - V G_0 Q)}^{-1} V \, ,
\end{equation}
which satisfies, as we expect, the standard equation for the optical potential
\begin{equation}\label{standard_optical_potential}
U = V + V G_0 Q U \, .
\end{equation}
After these formal manipulations we are ready to work out the left state vector of the transition amplitude.
With Eq.~(\ref{second_definition_elastic_opt_pot}) we can rewrite Eq.~(\ref{third_general_transition_amplitude}) as
\begin{equation}\label{fourth_general_transition_amplitude}
T_{f i}^{\mathrm{inel}} = \braket{\Psi_{\ast}^{(-)} | V | \psi^{(+)}} \, ,
\end{equation}
where $\bra{\Psi_{\ast}^{(-)}}$ satisfies
\begin{equation}\label{first_definition_of_final_dist_wave}
\bra{\Psi_{\ast}^{(-)}} = \bra{{\bm k}^{\prime} \Phi_{\ast}} + \bra{\Psi_{\ast}^{(-)}} V Q G_0 \, .
\end{equation}
If we now act with the operators $P_{\ast}$ and $Q_{\ast}$ on Eq.~(\ref{first_definition_of_final_dist_wave}) we obtain (considering that $Q P_{\ast} = P_{\ast}$)
\begin{equation}\label{first_elastic_final_dist_wave}
\bra{\Psi_{\ast}^{(-)}} P_{\ast} = \bra{{\bm k}^{\prime} \Phi_{\ast}} + \bra{\Psi_{\ast}^{(-)}} V P_{\ast} G_0 \, ,
\end{equation}
and
\begin{equation}\label{first_inelastic_final_dist_wave}
\bra{\Psi_{\ast}^{(-)}} Q_{\ast} = \bra{\Psi_{\ast}^{(-)}} V Q Q_{\ast} G_0 \, .
\end{equation}
Acting with the identity $\id = P_{\ast} + Q_{\ast}$ on the state vector $\bra{\Psi_{\ast}^{(-)}}$, on the right-hand side of Eq.~(\ref{first_elastic_final_dist_wave})
and Eq.(\ref{first_inelastic_final_dist_wave}), we obtain
\begin{equation}\label{second_elastic_final_dist_wave}
\bra{\Psi_{\ast}^{(-)}} P_{\ast} = \bra{{\bm k}^{\prime} \Phi_{\ast}} + \bra{\Psi_{\ast}^{(-)}} (P_{\ast} + Q_{\ast}) V P_{\ast} G_0 \, ,
\end{equation}
and
\begin{equation}\label{second_inelastic_final_dist_wave}
\bra{\Psi_{\ast}^{(-)}} Q_{\ast} = \bra{\Psi_{\ast}^{(-)}} P_{\ast} V Q Q_{\ast} G_0 {(\id - V Q Q_{\ast} G_0)}^{-1} \, .
\end{equation}
Now we are ready to obtain a second auxiliary optical potential. Noting that
\begin{equation}
\bra{\Psi_{\ast}^{(-)}} = \bra{\Psi_{\ast}^{(-)}} P_{\ast} + \bra{\Psi_{\ast}^{(-)}} Q_{\ast} \, ,
\end{equation}
we combine it with Eq.~(\ref{second_inelastic_final_dist_wave}) to obtain
\begin{equation}\label{reexpression_of_final_distort_wave}
\bra{\Psi_{\ast}^{(-)}} = \bra{\Psi_{\ast}^{(-)}} P_{\ast} {(\id - V Q Q_{\ast} G_0)}^{-1} \, .
\end{equation}
Inserting Eq.~(\ref{reexpression_of_final_distort_wave}) into Eq.~(\ref{fourth_general_transition_amplitude}), and defining
\begin{equation}\label{first_definition_final_potential}
\widehat{U} \equiv {(\id - V Q Q_{\ast} G_0)}^{-1} V \, ,
\end{equation}
and
\begin{equation}\label{definition_final_distorted_wave}
\bra{\psi_{\ast}^{(-)}} \equiv \bra{\Psi_{\ast}^{(-)}} P_{\ast} \, ,
\end{equation}
with the property
\begin{equation}\label{property_final_dist_wave}
\bra{\psi_{\ast}^{(-)}} = \bra{\psi_{\ast}^{(-)}} P_{\ast} \, ,
\end{equation}
the transition matrix element for inelastic scattering of Eq.~(\ref{fourth_general_transition_amplitude}) becomes
\begin{equation}\label{dwa_transition_matrix_element}
T_{f i}^{\mathrm{inel}} = \braket{\psi_{\ast}^{(-)}|P_{\ast} \widehat{U} P|\psi^{(+)}} \, .
\end{equation}
This expression is very similar to the conventional formulation of the DW formalism \cite{Satchler:1987ne}.
If we now insert Eq.~(\ref{second_inelastic_final_dist_wave}) into Eq.~(\ref{second_elastic_final_dist_wave}) and we use Eqs.~(\ref{first_definition_final_potential}),
(\ref{definition_final_distorted_wave}) and (\ref{property_final_dist_wave}), we obtain the equation for the final distorted wave $\bra{\psi_{\ast}^{(-)}}$
\begin{equation}\label{final_distorted_wave}
\bra{\psi_{\ast}^{(-)}} = \bra{{\bm k}^{\prime} \Phi_{\ast}} + \bra{\psi_{\ast}^{(-)}} P_{\ast} \widehat{U} P_{\ast} G_0 \, ,
\end{equation}
where the complete  distorting potential $\widehat{U}$ defined in Eq.~(\ref{first_definition_final_potential}) satisfies
\begin{equation}\label{distorting_potential}
\widehat{U} = V + V G_0 Q Q_{\ast} \widehat{U} \, .
\end{equation}
We immediately see that Eq.~(\ref{standard_optical_potential}) is the usual formal definition of the microscopic optical potential operator and
$P U P$ is the elastic optical potential that we already employed in our previous elastic scattering calculations \cite{Vorabbi:2015nra,Vorabbi:2017rvk,Vorabbi:2018bav,PhysRevC.103.024604,Vorabbi:2021kho} and from which one may calculate $\ket{\psi^{(+)}}$.
We note from Eq.~(\ref{dwa_transition_matrix_element}) and from Eq. (\ref{final_distorted_wave}) that the distorting potential which generates the final distorted wave,
{\it i.e.}, $P_{\ast} \widehat{U} P_{\ast}$, can be obtained from the diagonal matrix element (with respect to the excited state $\Phi_{\ast}$ of the target) of the same
operator $\widehat{U}$ whose off-diagonal matrix element, $P_{\ast} \widehat{U} P$, describes the inelastic transition.
The matrix element $T_{f i}^{\mathrm{inel}}$ can be calculated once the three effective interactions $P U P$, $P_{\ast} \widehat{U} P_{\ast}$, and $P_{\ast} \widehat{U} P$
are known. These effective interactions can be distinguished in:
\begin{enumerate}
\item the initial distorting potential 
\begin{equation} 
U_{\mathrm{el}} \equiv P U P \;,
\end{equation}
\item the final distorting potential
\begin{equation}
U_{\mathrm{ex}} \equiv P_{\ast} \widehat{U} P_{\ast} \;,
\end{equation}
\item the transition potential 
\begin{equation}
U_{\mathrm{tr}} \equiv P_{\ast} \widehat{U} P \;.
\end{equation}
\end{enumerate}

It is important to note that the potential $P_{\ast} \widehat{U} P_{\ast}$ does not correspond to the optical potential describing elastic scattering from the target in its
excited state because the operator $\widehat{U}$ has both the excited state $\ket{\Phi_{\ast}}$ and the ground state $\ket{\Phi}$ removed from its intermediate states due
to the presence of the operators $Q$ and $Q_{\ast}$. A proper optical potential for an excited state would only have the excited state $\ket{\Phi_{\ast}}$ projected out.

We conclude this section reminding that the transition matrix element of Eq.~(\ref{dwa_transition_matrix_element}) has been obtained from
Eq.~(\ref{definition_of_transition_amplitude}) without any approximation up to this point.

\subsection{Algorithm}
\label{algorithm}

A proposed scheme for a computational approach to the problem can be summarized as follows.
\begin{enumerate}
\item Employ Eq.~(\ref{standard_optical_potential}) for the calculation of the elastic optical potential $PUP$ at the first order of the Watson multiple scattering
theory and adopting the impulse approximation.
\item Employ Eq.~(\ref{initial_distorted_wave}) to calculate the initial distorted wave $\ket{\psi^{(+)}}$.
\item Employ Eq.~(\ref{distorting_potential}) for the calculation of the potentials $P_{\ast} \widehat{U} P$ and $P_{\ast} \widehat{U} P_{\ast}$ adopting the same
approximations used to calculate the potential $PUP$ in point 1.
\item Employ Eq.~(\ref{final_distorted_wave}) to calculate the final distorted wave $\bra{\psi_{\ast}^{(-)}}$.
\item Collect all previous calculations and calculate Eq.~(\ref{dwa_transition_matrix_element}) for the inelastic transition amplitude.
\end{enumerate}
In App. \ref{computation} we describe in detail every computational step of our algorithm.

\subsection{Distorted wave impulse approximation}
In this short subsection we show some practical approximations such that the previous equations can be actually calculated.
We already know that the operator $U$, Eq.~(\ref{standard_optical_potential}), can be expressed within the spectator expansion as
\begin{equation}
U = \sum_{i=1}^A \tau_i + \sum_{i,j\neq i}^A \tau_{ij} + \sum_{i,j\neq i,k\neq i,j}^A \tau_{ijk} + \cdots \, ,
\end{equation}
where
\begin{equation}\label{first_order_tau_oper}
\tau_i = v_{0i} + v_{0i} G_0 Q \tau_i \, .
\end{equation}
As we did in our previous works \cite{Vorabbi:2015nra,Vorabbi:2017rvk,Vorabbi:2018bav,PhysRevC.103.024604,Vorabbi:2021kho}, we approximate the operator $U$
with its first term, and we treat $\tau_i$ as explained in Ref. \cite{PhysRevC.52.1992}.
Assuming the impulse approximation, we approximate $\tau_i$ with the free $NN$ scattering matrix which satisfies
\begin{equation}
t_{0i} = v_{0i} + v_{0i} g_i t_{0i} \, ,
\end{equation}
with the free two-nucleon propagator
\begin{equation}
g_i = \frac{1}{(E-E^i) - h_0 - h_i + i \epsilon} \, .
\end{equation}
Following the same arguments, we expand the operator $\widehat{U}$, Eq.~(\ref{distorting_potential}), in a similar way
\begin{equation}
\widehat{U} = \sum_{i=1}^A \widehat{\tau}_i + \sum_{i,j\neq i}^A \widehat{\tau}_{ij} + \sum_{i,j\neq i,k\neq i,j}^A \widehat{\tau}_{ijk} + \cdots \, ,
\end{equation}
where
\begin{equation}\label{first_order_tauhat_oper}
\widehat{\tau}_i = v_{0i} + v_{0i} G_0 Q Q_{\ast} \widehat{\tau}_i \, .
\end{equation}
Also in this case, we approximate $\widehat{U}$ with its first term and, introducing the impulse approximation, we further approximate $\widehat{\tau}_i$ with $t_{0i}$.
With these approximations, the three potentials become
\begin{align}
U_{\mathrm{el}} &= \sum_{i=1}^A P t_{0i} P \, , \label{elastic_IA_potential} \\
U_{\mathrm{ex}} &= \sum_{i=1}^A P_{\ast} t_{0i} P_{\ast} \, , \\
U_{\mathrm{tr}} &= \sum_{i=1}^A P_{\ast} t_{0i} P \, . \label{transition_IA_potential}
\end{align}
The domain of applicability of this approximation, in particular concerning the energy of the incoming projectile, will be tested in Sec. \ref{results}.

\section{Derivation of the potentials}
\label{theory}

In the previous section we derived the Distorted Wave Impulse Approximation (DWIA) equations which require the knowledge of three potentials.
In this section we briefly mention how the three potentials are derived.
Defining with ${\bm k}$ and ${\bm k}^{\prime}$ the initial and final momenta of the projectile in the $NA$ reference frame, and using $\Phi_i$ and $\Phi_f$ to represent
the initial and final target state, the three potentials are obtained as
\begin{equation}
\begin{split}
U_{f i}^{\textbf{p}} ({\bm k}^{\prime} , {\bm k}) &= Z \braket{{\bm k}^{\prime} \Phi_f | t_{\textbf{p} p} (E) | \Phi_i {\bm k}} \\
&+ N \braket{{\bm k}^{\prime} \Phi_f | t_{\textbf{p} n} (E) | \Phi_i {\bm k}} \, ,
\end{split}
\end{equation}
where $Z$ and $N$ are the proton and neutron numbers, respectively, and $\textbf{p}$ denotes the incoming projectile (either a proton $p$ or a neutron $n$).
Due to the approximations introduced in the previous section, and labelling with $0$ the projectile nucleon and with $A$ the $A$-th nucleon in the target nucleus,
the three potentials given in Eqs.~(\ref{elastic_IA_potential})-(\ref{transition_IA_potential}) can be obtained evaluating the general matrix element
\begin{equation}
\braket{t_{f i}} \equiv \braket{{\bm k}^{\prime} \Phi_f | t_{0 A} (E) | \Phi_i {\bm k}}
\end{equation}
for different choices of $\Phi_i$ and $\Phi_f$.
For example, if we choose $\Phi_i = \Phi_f = \Phi$ we obtain the optical potential $U_{\mathrm{el}}$, whose derivation can be found in Ref.\cite{PhysRevC.85.044617}
and leads to the optical potential for the elastic scattering that we used in past
works \cite{Vorabbi:2015nra,Vorabbi:2017rvk,Vorabbi:2018bav,PhysRevC.103.024604,Vorabbi:2021kho}.
Additionally, if we choose $\Phi_i = \Phi_f = \Phi_{\ast}$ we obtain the potential $U_{\mathrm{ex}}$, while if we choose $\Phi_i = \Phi$ and
$\Phi_f = \Phi_{\ast}$ we obtain the potential $U_{\mathrm{tr}}$. In this way, we can simply generalize the derivation given in Ref.\cite{PhysRevC.85.044617} to obtain
the three potentials. The only thing that will change in the final formula for the three potentials is the one-body density matrix, that will be different depending on the
above choices of $\Phi_i$ and $\Phi_f$.
Apart from this difference, the derivation of the final expression for the three potentials is exactly the same.
Specifically, we have
\onecolumngrid
\begin{align}
U_{\mathrm{el}}^{\textbf p} ({\bm q} , {\bm K}) &= \sum_{\mathcal{N} =p,n} \int d {\bm P} \; \eta ({\bm q} , {\bm K} , {\bm P}) \;
t_{{\textbf p} \mathcal{N}} \left[ {\bm q} , \frac{1}{2} \left( \frac{A+1}{A} {\bm K} + \sqrt{\frac{A-1}{A}} {\bm P} \right) ; E \right] \nonumber \\
&\times \rho_{\mathcal{N}}^{\mathrm{gs}} \left( {\bm P} + \frac{1}{2} \sqrt{\frac{A-1}{A}} {\bm q} , {\bm P} - \frac{1}{2} \sqrt{\frac{A-1}{A}} {\bm q} \right) \, ,
\label{elastic_pot_final_expression} \\
U_{\mathrm{ex}}^{\textbf p} ({\bm q} , {\bm K}) &= \sum_{\mathcal{N} =p,n} \int d {\bm P} \; \eta ({\bm q} , {\bm K} , {\bm P}) \;
t_{{\textbf p} \mathcal{N}} \left[ {\bm q} , \frac{1}{2} \left( \frac{A+1}{A} {\bm K} + \sqrt{\frac{A-1}{A}} {\bm P} \right) ; E \right] \nonumber \\
&\times \rho_{\mathcal{N}}^{\mathrm{ex}} \left( {\bm P} + \frac{1}{2} \sqrt{\frac{A-1}{A}} {\bm q} , {\bm P} - \frac{1}{2} \sqrt{\frac{A-1}{A}} {\bm q} \right)  \, , \\
U_{\mathrm{tr}}^{\textbf p} ({\bm q} , {\bm K}) &= \sum_{\mathcal{N} =p,n} \int d {\bm P} \; \eta ({\bm q} , {\bm K} , {\bm P}) \;
t_{{\textbf p} \mathcal{N}} \left[ {\bm q} , \frac{1}{2} \left( \frac{A+1}{A} {\bm K} + \sqrt{\frac{A-1}{A}} {\bm P} \right) ; E \right] \nonumber \\
&\times \rho_{\mathcal{N}}^{\mathrm{tr}} \left( {\bm P} + \frac{1}{2} \sqrt{\frac{A-1}{A}} {\bm q} , {\bm P} - \frac{1}{2} \sqrt{\frac{A-1}{A}} {\bm q} \right) \, .
\label{transition_pot_final_expression}
\end{align}
\twocolumngrid
Here, ${\bm q} = {\bm k}^{\prime} - {\bm k}$ is the momentum transfer, ${\bm K} = ( {\bm k}^{\prime} + {\bm k} ) / 2$ is the average momentum, ${\bm P}$ is
an integration variable, and $\eta$ is the M\o ller factor that imposes the Lorentz invariance when the $t$ matrix is transformed from the $NN$ frame, where it is evaluated,
to the $NA$ frame, where it is needed for the calculation of the potential.

The three formulas above only differ by the presence of a different density. Here, we used $\rho_{\mathcal{N}}^{\mathrm{gs}}$ to indicate
the neutron and proton ground-state density, $\rho_{\mathcal{N}}^{\mathrm{ex}}$ for the neutron and proton density of the excited state,
and $\rho_{\mathcal{N}}^{\mathrm{tr}}$ for the neutron and proton transition density.

In general, the three potentials depend also on the initial and final third components, $\sigma$ and $\sigma^{\prime}$, of the target spin,
that are omitted here for the sake of brevity. More details about the inclusion of non-zero spin targets can be found in Ref. \cite{PhysRevC.103.024604}.
All the details about the calculation of the density can be found in Sect.~\ref{target}.
Finally, the energy $E$ in the $NN$ $t$ matrix displays a dependence on the integration variable ${\bm P}$ and makes the calculation of the integral very complicated.
In our calculations we assume the so called fixed beam energy approximation, which consists in setting $E$ equal to one-half the kinetic energy of the projectile in the
laboratory frame.

\section{Nucleon-nucleon interaction}
\label{NNpotential}

The theoretical framework we use to define the $NN$ interaction is Chiral Perturbation Theory (ChPT).
It is a perturbative technique for the description of hadron scattering amplitudes based on expansions in powers of a parameter that can be generally defined
as $(p,m_\pi)/\Lambda_b$, where $p$ is the magnitude of 3-momenta of the external particles, $m_\pi$ is the pion mass, and the symmetry breaking
scale $\Lambda_b$ can be safely estimated for chiral symmetry as follows $\Lambda_b \sim 4 \pi f_\pi$ \cite{Scherer:2002tk} or, alternatively, using the
lightest non-Goldstone meson mass as an energy scale, $\Lambda_b \sim m_\rho$. 

As an EFT, ChPT respects the low-energy symmetries of QCD and, up to a certain extent, is model independent and systematically improvable by an order-by-order
expansion, with controlled uncertainties from neglected higher-order terms.

Among the different versions of $NN$ potentials derived from ChPT we have chosen the approach developed by Entem {\it et al.} \cite{Entem:2017gor}.
Calculations are performed up to the fifth order (N$^{4}$LO) with a 500 MeV cutoff and the three-nucleon ($3N$) local-nonlocal chiral interaction at N$^{2}$LO presented
in Refs. \cite{Navratil:2007we, Navratil:2007zn,PhysRevC.101.014318}, with the low-energy constants fixed at $c_D = -1.8$ and $c_E =-0.31$ \cite{Gysbers2019}.
The same $NN$ chiral interaction used to compute the nuclear densities is consistently used in the calculation of the $NN$ $t$ matrix.

\section{Target description}
\label{target}

The three nonlocal densities $\rho^{\mathrm{gs}}$, $\rho^{\mathrm{ex}}$, and $\rho^{\mathrm{tr}}$ are an important ingredient of the calculation, and they
are obtained employing the NCSM method~\cite{Navr_til_2009,Barrett:2013nh}. 
This approach is based on the expansion of the nuclear wave function in a harmonic oscillator basis and it is thus characterized by the harmonic
oscillator frequency $\hbar \omega$ and the $N_{max}$ parameter, which specifies the number of nucleon excitations above the lowest energy configuration
allowed by the Pauli principle.
For the nucleus considered in this work we used $N_{max} = 8$ excitations and $\hbar \omega = 16$ MeV.
The $NN$ and $3N$ interactions have been also softened using the Similarity Renormalization Group (SRG) \cite{bogner_2007} procedure using a
$\lambda_{\mathrm{SRG}} = 2.0$ fm$^{-1}$ cutoff, and including the SRG induced $3N$ force in all calculations.


\section{Results}
\label{results}

In this section we test the reliability of our microscopic model for inelastic proton scattering in comparison with experimental differential cross sections off $^{12}$C leading to the first $2^+$ state at 4.44 MeV in a range of proton energies between 65 and 300 MeV. 

We point out that our model requires as only input the chiral interactions needed for the calculations of the nuclear densities and of the $NN$ $t$-matrix (see Sec. \ref{target} and Sec. \ref{NNpotential}). No free parameters nor ad hoc adjustments have been introduced to reproduce empirical data. 

In order to avoid any misleading interpretation, we dropped the usual correction introduced in most DWBA/DWIA calculations, where the transition densities are
scaled in order to have a correct $BE(2)$.

\begin{figure}[t]
\begin{center}
\includegraphics[angle=0, scale=0.5]{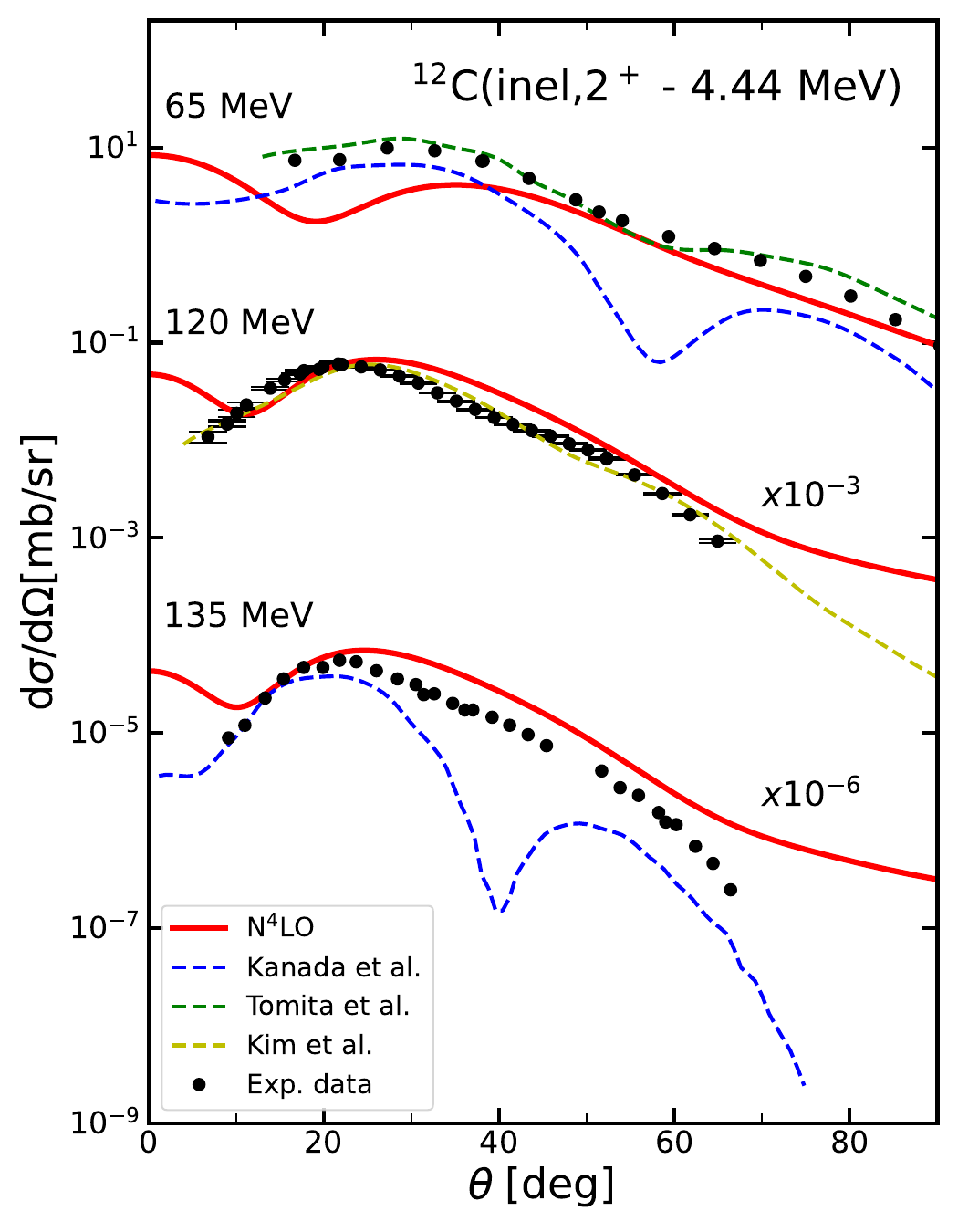}
\caption{ \label{fig1} 
Differential cross sections as functions of the scattering angle for inelastic proton scattering off $^{12}$C leading to the $2^+$ state at 4.44 MeV for different projectile energies: 65, 120 ($\times 10^{-3}$), 135 MeV ($\times 10^{-6}$). The experimental data are taken from Refs. \cite{Pignanelli:1986zz,PhysRevC.31.1616,PhysRevC.24.1834,BAUHOFF1983180}. The red lines show the results of our microscopic model. The results of other available theoretical models are also shown for a comparison: DWBA calculations with the AMD+GCM densities \cite{PhysRevC.100.064616} (dashed blue line), $g$-folding model with NCSM densities \cite{Kim:2007cg} (dashed yellow line), and phenomenological coupled-channel calculation with the $3\alpha$-RGM densities and $DD3MY$ interaction \cite{PhysRevC.92.024609} (dashed green line).}
\end{center}
\end{figure}

\begin{figure}[t]
\begin{center}
\includegraphics[angle=0, scale=0.5]{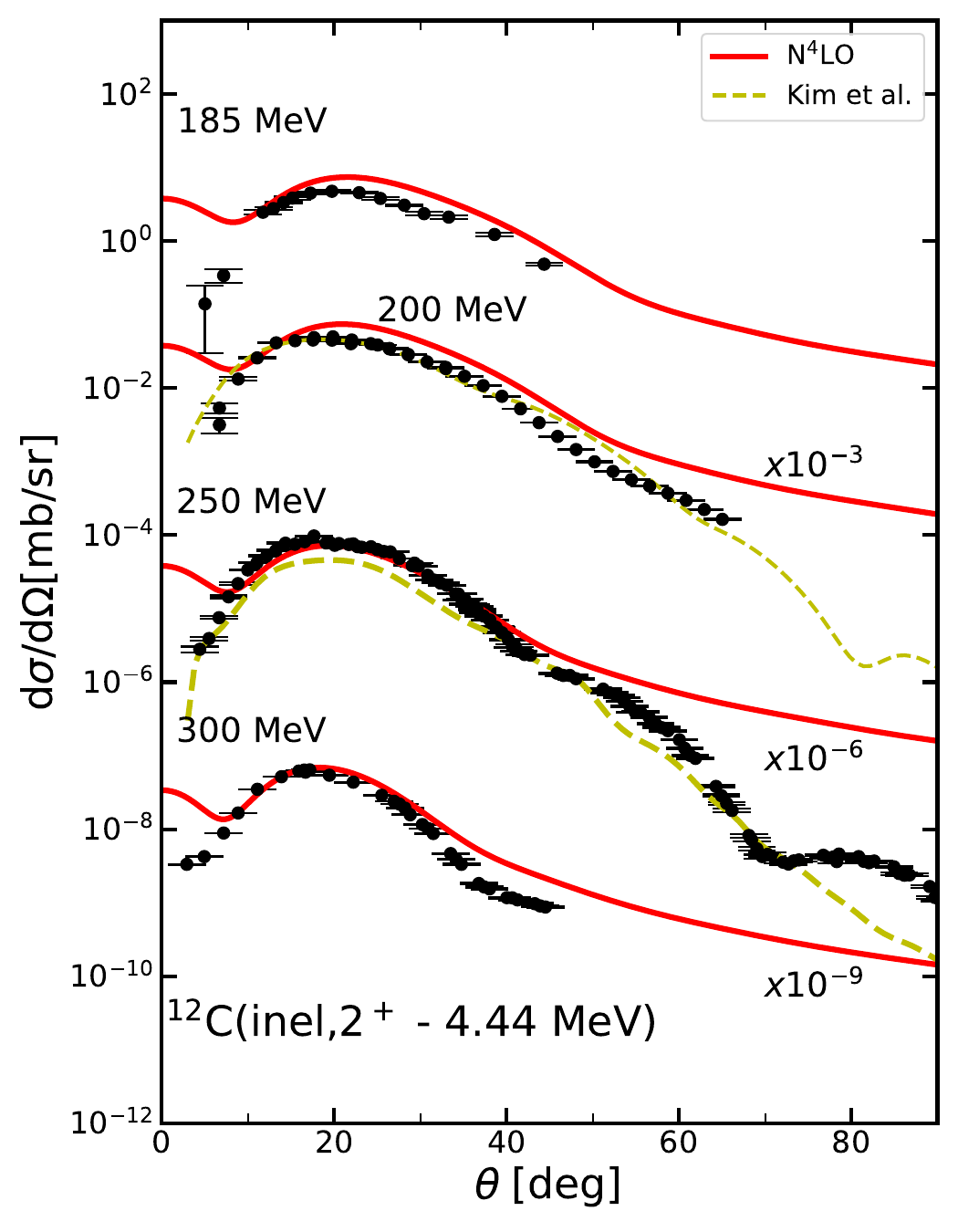}
\caption{ \label{fig2} 
Same as in Fig. \ref{fig1} but for the projectile energies of 185, 200($\times 10^{-3}$), 250 ($\times 10^{-6}$), and  300 MeV ($\times 10^{-9}$). The experimental data are taken from Refs. \cite{Ingemarsson:1979eb,PhysRevC.26.1800,PhysRevC.37.544,Okamoto:2010zzb}.}
\end{center}
\end{figure}

Our theoretical predictions are plotted in Figs. \ref{fig1} and \ref{fig2} with continuous red lines. Besides experimental data \cite{Pignanelli:1986zz,PhysRevC.31.1616,PhysRevC.24.1834,BAUHOFF1983180,Ingemarsson:1979eb,PhysRevC.26.1800,PhysRevC.37.544,Okamoto:2010zzb}, which are available in a wide energy range, we also display the results of other theoretical models available in the literature as benchmarks.

In Ref. \cite{Kim:2007cg} (Kim {\it et al.}), the authors have studied this particular inelastic transition for a large energy range (from 35 to 250 MeV), employing a microscopically inspired approach.
They determined optical potentials using a $g$-folding model and analyzed the inelastic data with a distorted-wave approximation.
In a consistent manner, as we have done, the effective $NN$ interactions used to specify the optical potentials have also been used as the transition
operators. The nuclear states have been described by a NCSM approach, but with a phenomenological interaction. 
The results of this approach (shown by the dashed yellow line in the figures) are able to give an excellent description of the experimental differential cross sections, in particular, for proton energies larger than 100 MeV. However, despite the undoubted quality, it lacks the overall consistency and the theoretical rigor that are the main strengths of our approach.

In Ref. \cite{PhysRevC.92.024609} (Tomita {\it et al.})  microscopic coupled-channel calculations have been performed for projectile energies between 30 and 65 MeV.
The nuclear interactions for the p+$^{12}$C system are constructed by a folding procedure, which employs the internal wave function of $^{12}$C,
obtained from the 3$\alpha$ resonating group method, and an effective $NN$ interaction of the density-dependent Michigan three-range Yukawa ($DDM3Y$).
The results of this approach, shown by a dashed green line only for a proton energy of 65 MeV in Fig. \ref{fig1}, give an excellent description of the experimental data.

In Ref. \cite{PhysRevC.100.064616} (Kanada {\it et al.}) microscopic coupled-channel calculations have been performed by folding the
Melbourne $g$-matrix $NN$ interaction with the matter and transition densities of the target nuclei obtained via the Antisymmetrized Molecular Dynamics (AMD) and
the generator coordinate method (GCM). The results of this approach, shown by the blue dashed lines in Fig. \ref{fig1}, give a reasonably good description of the experimental cross sections at forward angles, while increasing the scattering angle the experimental cross section is significantly underestimated. 

The results of our microscopic approach, based on the distorted-wave approximation and on chiral EFTs interactions at N$^4$LO, are remarkably comparable with experimental data for proton energies above 100 MeV. For lower energies the magnitude of the experimental differential cross section is somewhat underestimated. This is presumably a consequence of the breakdown of the impulse approximation. Extension of our framework to DWBA or, even better, to coupled-channel calculations, is a natural improvement for the future that should reduce the discrepancies with data.
 
Our results are in excellent agreement with the measured cross section across several energies, in a large energy range.
This underscores the robustness of the chiral EFT interactions used and highlights the predictive power of the distorted-wave approach when applied consistently across different kinematic regimes.
Notably, the magnitude of the cross section is well reproduced without the need for any empirical normalization. This indicates that both the transition operator and the nuclear structure input, such as the transition densities, are reliably described within the chosen framework.
The angular distributions are also accurately captured, with the calculations closely following the diffraction patterns observed in the data, including the positions and
depths of minima and maxima. This level of agreement suggests that nuclear distortion effects, encoded in the distorted incoming and outgoing waves, are realistically
accounted for by the optical model potentials employed. 

The disagreement between our results and data at small and large angles, where the calculated cross sections generally overestimate the measured ones, deserves further investigations and requires improvements in the model, which has some merits but contains several approximations. Improvements are required to extend the reliability of the model to a wider range of situations. For instance, the potentials derived and used in our approach contain only central and spin-orbit terms, while other neglected terms, like the tensor term, might be significant for the transition potential.

Till now the model has been tested only for the differential cross sections of a given nucleus and a specific transition. The reliability of our framework remains to be tested for other nuclei, other nuclear states and polarization observables such as the analyzing power. Work along this direction is currently in progress.
 

\section{Conclusions}
\label{conclusions}

We have investigated the inelastic scattering of protons off $^{12}\text{C}$ leading to the excitation of the first $2^+$ state at 4.44 MeV by employing a fully
microscopic reaction framework based on the distorted-wave impulse approximation. The underlying nuclear interactions and transition operators were derived from chiral
effective field theory at fifth order (N$^4$LO), ensuring consistency between the structure and reaction components of the calculation.

A central feature of the analysis is the use of distorted waves, generated from global microscopic potentials tailored to the specific energies under consideration,
which provide a realistic description of initial- and final-state interactions, going beyond the plane-wave approximation and incorporating elastic channel effects in a
non-perturbative way. The nuclear densities used in the calculation of the potentials are all obtained from the NCSM method and are consistently linked to the chiral
interaction employed in the reaction operator. The only input to the model is thus the chiral nuclear force. No adjustable free
parameters or phenomenological normalizations are introduced.

The resulting differential cross sections show excellent agreement with experimental data across a wide angular range and at most incident energies (above
$100$ MeV of the projectile energy). The calculations not only reproduce the magnitude of the measured cross sections but also accurately capture the detailed
diffraction pattern, in the region of the first maximum. This confirms that the long-range and medium-range components of the nuclear
interaction, as encoded in the chiral EFT potential, are realistically represented, and that the distorted-wave approach accurately incorporates the relevant reaction dynamics.

Furthermore, the observed agreement is stable with respect to variations in energy, which demonstrates the robustness and scalability of the approach.
The quality of the results suggests that the combination of high-order chiral EFT and the DWIA formalism can serve as a reliable tool for the description of inelastic
scattering involving medium-mass nuclei, at both intermediate and high energies. 

Comparison with other theoretical results from the literature indicates that the present 
calculations offer systematically improved predictive power, particularly at large angles and for higher momentum transfers.

These findings highlight the strength of a fully microscopic, theoretically grounded reaction framework that avoids the ambiguities inherent in phenomenological fitting.
They also open the path toward applications to more complex reactions, including transfer and knockout processes, and to nuclei further from stability.
Future work will focus on extending the method to include multistep and coupled-channel effects within a consistent EFT framework.


\section*{Acknowledgements}

This work used the DiRAC Data Intensive service (DIaL3) at the University of Leicester, managed by the University of Leicester Research Computing Service on
behalf of the STFC DiRAC HPC Facility (www.dirac.ac.uk). The DiRAC service at Leicester was funded by BEIS, UKRI and STFC capital funding and STFC operations
grants. DiRAC is part of the UKRI Digital Research Infrastructure. This work used the DiRAC Complexity system, operated by the University of Leicester IT Services,
which forms part of the STFC DiRAC HPC Facility (www.dirac.ac.uk ). This equipment is funded by BIS National E-Infrastructure capital grant ST/K000373/1 and STFC
DiRAC Operations grant ST/K0003259/1. DiRAC is part of the National e-Infrastructure.
This work was supported in part by the Deutsche Forschungsgemeinschaft (DFG) through the Cluster of Excellence "Precision Physics, Fundamental Interactions,
and Structure of Matter'" (Project ID 390831469).
P. N. acknowledges support from the NSERC Grant No. SAPIN-2022-00019. TRIUMF receives federal funding via a contribution agreement with the National
Research Council of Canada. Computing support also came from an INCITE Award on the Summit and Frontier supercomputers of the Oak Ridge Leadership Computing
Facility (OLCF) at ORNL and from the Digital Research Alliance of Canada.


\begin{appendix}

\section{Computational framework}
\label{computation}

In this appendix we want to explore the algorithm proposed in Sect. \ref{algorithm}  to describe inelastic scattering processes.
The central goal is to compute the inelastic transition amplitude of Eq.~(\ref{dwa_transition_matrix_element}), that with the previous definitions can be written as
\begin{equation}
T_{f i}^{\mathrm{inel}} = \braket{ \psi_{\ast}^{(-)} | U_{\mathrm{tr}} | \psi^{(+)} } \, .
\end{equation}
For this we need the transition potential $U_{\mathrm{tr}}$, describing the inelastic excitation of the target, and the two distorted waves
\begin{align}
\ket{\psi^{(+)}} &= \ket{\Phi \, {\bm k}} + G_0 U_{\mathrm{el}} \ket{\psi^{(+)}} \, , \label{initial_distorted_wave} \\
\bra{\psi_{\ast}^{(-)}} &= \bra{{\bm k}^{\prime} \Phi_{\ast}} + \bra{\psi_{\ast}^{(-)}} U_{\mathrm{ex}} G_0 \, . \label{final_distorted_wave}
\end{align}
As a consequence, the inputs to calculate the previous quantities are the three potentials $U_{\mathrm{el}}$, $U_{\mathrm{ex}}$, and $U_{\mathrm{tr}}$,
given in Eqs.~(\ref{elastic_pot_final_expression})-(\ref{transition_pot_final_expression}).
To avoid numerical issues related to the treatment of the delta function leading term in Eq.~(\ref{initial_distorted_wave}) and Eq.~(\ref{final_distorted_wave}) in
momentum space, we propose a solution in the coordinate space.
Introducing a resolution of the identity,
we can express the inelastic transition amplitude as
\onecolumngrid
\begin{equation}\label{inel_trans_amp_coordinate_space}
T_{\nu^{\prime} s^{\prime} \sigma^{\prime} \nu s \sigma}^{\mathrm{inel}} ( {\bm k}_{\ast} , {\bm k}_0 ; E) = \int d {\bm r}^{\prime} d {\bm r} \;
\psi_{\ast}^{(-) \, \dagger} ({\bm k}_{\ast} \nu^{\prime} s^{\prime} \sigma^{\prime} ; {\bm r}^{\prime}) 
U_{\mathrm{tr}} ({\bm r}^{\prime} , {\bm r} ; s^{\prime} \sigma^{\prime} s \, \sigma E) 
\psi^{(+)} ({\bm k}_0 \nu s \sigma ; {\bm r}) \, .
\end{equation}
Here, we used the multi-index $\nu$ and $\nu^{\prime}$ to indicate the initial and final projectile spin polarizations, we used $s$ and $s^{\prime}$
to label the initial and final target spin, and $\sigma$ and $\sigma^{\prime}$ their corresponding polarizations.
The momenta ${\bm k}_0$ and ${\bm k}_{\ast}$ represent the initial
and final momenta of the projectile nucleon in the $NA$ center of mass, respectively.

We expand the potentials in coordinate and momentum space as
\begin{align}
U ({\bm r}^{\prime} , {\bm r} ; s^{\prime} \sigma^{\prime} s \, \sigma E) &= \sum_{l j m} \mathcal{Y}_{j m}^{l {\scriptstyle \frac{1}{2}}} (\hat{\bm r}^{\prime}) \,
U_{l j} (r^{\prime} , r ; s^{\prime} \sigma^{\prime} s \, \sigma E) \,
\mathcal{Y}_{j m}^{l {\scriptstyle \frac{1}{2}} \, \dagger} (\hat{\bm r}) \, , \label{potential_in_coordinate_space} \\
U ({\bm k}^{\prime} , {\bm k} ; s^{\prime} \sigma^{\prime} s \, \sigma E) &= \frac{2}{\pi} \sum_{l j m} \mathcal{Y}_{j m}^{l {\scriptstyle \frac{1}{2}}} (\hat{\bm k}^{\prime}) \,
U_{l j} (k^{\prime} , k ; s^{\prime} \sigma^{\prime} s \, \sigma E) \, \mathcal{Y}_{j m}^{l {\scriptstyle \frac{1}{2}} \, \dagger} (\hat{\bm k}) \, , \label{potential_in_momentum_space}
\end{align}
where $\mathcal{Y}_{j m}^{l {\scriptstyle \frac{1}{2}}}$ are the standard spin-angular functions.
The partial wave components of the potential in coordinate space can be obtained from those in momentum space as
\begin{equation}
U_{l j} (r^{\prime} , r ; s^{\prime} \sigma^{\prime} s \, \sigma E) \equiv \frac{4}{\pi^2} \int_0^{\infty} d k^{\prime} k^{\prime \, 2} \int_0^{\infty} d k k^2 \, j_l (k^{\prime} r^{\prime}) \,
U_{l j} (k^{\prime} , k ; s^{\prime} \sigma^{\prime} s \, \sigma E) \, j_l (k r) \, ,
\end{equation}
where $j_l (k r)$ are spherical Bessel functions.

We expand the distorted wave function as
\begin{equation}\label{expansion_distorted_wave}
\begin{split}
\psi ({\bm k}_0 \nu s \sigma ; {\bm r}) = \sqrt{\frac{2}{\pi }} \sum_{l j m} i^l e^{i \sigma_l} \frac{\psi_{l j} (k_0 s \sigma ; r)}{k_0 r} \,
\mathcal{Y}_{j m}^{l {\scriptstyle \frac{1}{2}} } (\hat{\bm r}) \, \mathscr{Y}_{j m}^{l {\scriptstyle \frac{1}{2}} \, \ast} (\nu ; \hat{\bm k}_0) \, ,
\end{split}
\end{equation}
where
\begin{equation}
\mathscr{Y}_{j m}^{l {\scriptstyle \frac{1}{2}}} (\nu ; \hat{\bm k}_0) \equiv \sum_{\lambda} (l \lambda {\scriptstyle \frac{1}{2}} \nu | j m) Y_{l \lambda} (\hat{\bm k}_0) \, .
\end{equation}

If we insert the expansions of the potentials and the distorted waves into Eq.~(\ref{inel_trans_amp_coordinate_space}) we obtain the partial wave components of the inelastic
transition amplitude
\begin{equation}\label{transition_amplitude_pw}
T_{l j}^{\mathrm{inel}} (k_{\ast} , k_0 ; s^{\prime} \sigma^{\prime} s \, \sigma E) = \frac{1}{k_{\ast} k_0} \int_0^{\infty} dr^{\prime} r^{\prime} \int_0^{\infty} dr r \;
\psi_{l j}^{\ast} (k_{\ast} s^{\prime} \sigma^{\prime} ; r^{\prime}) \, U_{l j} (r^{\prime} , r ; s^{\prime} \sigma^{\prime} s \, \sigma E) \, \psi_{l j} (k_0 s \sigma ; r) \, .
\end{equation}
\twocolumngrid
The scattering amplitude for the inelastic excitation of a target from a state with spin $s$ to a state with spin $s^{\prime}$ is given by
\begin{equation}\label{general_inelastic_scattering_amplitude}
\begin{split}
f_{\nu^{\prime} \sigma^{\prime} \nu \, \sigma}^{\mathrm{inel}} (\theta) &= \frac{2}{\pi} \sum_{l^{\prime} j^{\prime} l j J \pi} (l 0 {\scriptstyle \frac{1}{2}} \nu | j \nu) \,
(j \nu s \sigma | J , \nu + \sigma) \\
&\times (l^{\prime} , \nu + \sigma - \nu^{\prime} - \sigma^{\prime}, {\scriptstyle \frac{1}{2}} \nu^{\prime} | j^{\prime} , \nu + \sigma - \sigma^{\prime}) \\
&\times (j^{\prime} , \nu + \sigma - \sigma^{\prime} , s^{\prime} \sigma^{\prime} | J , \nu + \sigma) \\
&\times e^{i [ \sigma_{l^{\prime}} (\eta(k_{\ast}) ) + \sigma_l ( \eta (k_0) ) ]} \sqrt{\frac{2 l + 1}{4 \pi}} \\
&\times M_{l^{\prime} j^{\prime} l j} (s^{\prime} \sigma^{\prime} s \, \sigma E) \\
&\times Y_{l^{\prime},\nu+\sigma - \nu^{\prime} - \sigma^{\prime}} (\theta , 0) \, ,
\end{split}
\end{equation}
where $\eta$ is the Sommerfeld parameter, $\sigma_l$ are the Coulomb phase shifts, and the partial wave components of the scattering amplitude are
obtained from the inelastic transition amplitude as
\begin{equation}
M_{l^{\prime} j^{\prime} l j} (s^{\prime} \sigma^{\prime} s \, \sigma E) = - 4 \pi^2 \mu \, \delta_{l^{\prime} l} \delta_{j^{\prime} j}
T_{l j}^{\mathrm{inel}} (k_{\ast} , k_0 ; s^{\prime} \sigma^{\prime} s \, \sigma E)  \, .
\end{equation}
From Eq.(\ref{general_inelastic_scattering_amplitude}) one can calculate the differential cross section 
employing the commonly used relation in the DWIA approach \cite{17293}
\begin{equation}\label{general_differential_cross_section_formula}
\frac{d \sigma}{d \Omega} ( \theta ) = \frac{2s^\prime + 1}{2 (2 s + 1)} \sum_{\nu^{\prime} \sigma^{\prime} \nu \, \sigma}
{\Big| f_{\nu^{\prime} \sigma^{\prime} \nu \, \sigma}^{\mathrm{inel}} (\theta) \Big|}^2 \, .
\end{equation}

Finally, the partial wave components of the distorted waves are obtained solving the Schr\"odinger equation with a short-range nonlocal nuclear potential
and a Coulomb potential. With these two potentials the Schr\"odinger equation becomes
\begin{equation}\label{schrodinger_eq_three_dim}
\begin{split}
&- \frac{\hbar^2 \nabla^2}{2 \mu} \psi ({\bm k} \nu s \sigma ; {\bm r}) + V_c (r) \, \psi ({\bm k} \nu s \sigma ; {\bm r}) \\
&+ \int d{\bm r}^{\prime} \, U ( {\bm r} , {\bm r}^{\prime} ; s \, \sigma E)  \, \psi ({\bm k} \nu s \sigma ; {\bm r}^{\prime}) \\
&= E \, \psi ({\bm k} \nu s \sigma ; {\bm r}) \, ,
\end{split}
\end{equation}
where ${\bm k}$ stands for either ${\bm k}_0$ or ${\bm k}_{\ast}$, and
\begin{equation}
V_c (r) =
\begin{cases}
\frac{Z z e^2}{2 R_c} \left( 3 - \frac{r^2}{R_c^2} \right), & \text{for} \: r \leq R_c \, , \\
\frac{Z z e^2}{r}, & \text{for} \: r \geq R_c \, ,
\end{cases}
\end{equation}
with $R_c = r_c A^{1/3}$.
Depending on the value of $s$ we can have the optical potential for the ground state or the potential for the excited state under consideration.
Inserting Eq.(\ref{potential_in_coordinate_space}) and Eq.(\ref{expansion_distorted_wave}) into Eq.(\ref{schrodinger_eq_three_dim}),
the Schr\"odinger equation becomes
\begin{equation}\label{schrodinger_equation_pw}
\begin{split}
&\left[ - \frac{\hbar^2}{2 \mu} \left( \frac{d^2}{d r^2} - \frac{l (l+1)}{r^2} \right) + V_c (r) \right] \psi_{l j} (k s \sigma ; r) \\
&+ \int_0^{\infty} dr^{\prime} W_{l j} (r,r^{\prime} ; s \sigma E) \, \psi_{l j} (k s \sigma ; r^{\prime}) \\
&= E \psi_{l j} (k s \sigma ; r) \, ,
\end{split}
\end{equation}
where $W_{l j} (r,r^{\prime} ; s \sigma E) = U_{l j} (r,r^{\prime} ; s \, \sigma \, s \, \sigma E) \,  r \, r^{\prime}$.
Eq.(\ref{schrodinger_equation_pw}) is solved for all the partial waves using the R matrix \cite{Descouvemont:2010cx} theory, to obtain the partial wave components
of the distorted waves that are used into Eq.(\ref{transition_amplitude_pw}) for the transition amplitudes.

\end{appendix}

%

\end{document}